# Determination of the characteristics of a linear ducted sound source

Timothy J. Newman, Anurag Agarwal and Ann P. Dowling

Department of Engineering, University of Cambridge, Trumpington Street, Cambridge, CB2 1PZ

Ludovic Desvard

Dyson Ltd., Tetbury Hill, Malmesbury, Wiltshire, SN16 0RP

**Abstract**

Ducted flow devices for a range of purposes, such as air-moving fans, are routinely characterised experimentally to understand their acoustic performance as part of the continuing trend for quiet, high efficiency design. The International Organization for Standardization (ISO) method 5136 is widely used in industry and academia to determine the sound radiated into a duct. This involves placing the device at the centre of a long cylindrical duct with anechoic terminations at each end to eliminate reflections. A single off-axis microphone is used on the inlet and outlet sides that can theoretically capture the plane-wave mode amplitudes but this does not provide enough information to fully account for higher-order modes. In this study, the 'two-port' source model is formulated to include higher-order modes and applied for the first three modes as a proof of concept. This requires six independent surface pressure measurements on each side or 'port'. The resulting experimental set-up is much shorter than the ISO rig and does not require anechoic terminations. The relative importance of the higher-order modes has been considered and the inaccuracies when using the ISO method to find source sound power has been analysed.

# LIST OF SYMBOLS

| | |
|---|---|
| $c_0$ | Speed of sound |
| $a$ | Duct radius |
| $p$ | Acoustic pressure |
| $v$ | Acoustic velocity |
| $P$ | Acoustic power |
| $J_m$ | Bessel functions of the first kind |
| $z_{m,n}$ | Bessel function derivative zeros |
| $m$ | Azimuthal mode number |
| $n$ | Radial mode number |
| $A_{m,n}^+$ | Amplitude mode travelling away from source |
| $A_{m,n}^-$ | Amplitude mode travelling towards source |
| $A_{m,n}^s$ | Amplitude mode travelling from source directly |
| $\hat{r}$ | Reference signal |
| $k$ | Wave number |
| $\hat{S}$ | Spectral density |
| $\hat{H}$ | Transfer function |
| $s$ | Cross-sectional area |
| $C_{m,n}$ | Normalisation factor |
| $M$ | Mach number |
| $\boldsymbol{M}$ | Mode decomposition matrix |
| $S$ | Scattering matrix |
| $\rho$ | Reflection coefficient |
| $\tau$ | Transmission coefficient |
| $\bar{I}_x$ | Time-averaged axial intensity |
| $\kappa$ | Condition number |
| $\lambda$ | Wavelength |
| $N$ | Number of propagating modes |
| SPL | Sound pressure level |
| PWL | Sound power level |
| $\varphi$ | Flow coefficient |
| $\psi$ | Non-dimensional pressure rise |

## I. INTRODUCTION

Many fluid-moving devices such as fans generate sound which radiates to the far field as a result of unsteadiness in the flow. In order to reach the listener, the sound waves often propagate through a system of ducts or flow passages. Measurement of the sound power emitted by an isolated device or the overall system is a common way to quantify the human response to noise exposure. This is widely done by industry and academia in a standardised way using an array of free-field measurements [1] or by measuring the sound emitted into a duct [2]. These methods take account of the sound field directionality by prescribing a range of conditions which involve temporal and spatial averages of multiple sound pressure level measurements. In this paper a more complete duct-based method is detailed to measure the underlying structure of the sound field. This could help us understand the characteristics of the source and the effects of the system in which it is installed.

The propagation of sound in a duct can be broken down into contributions from the one-dimensional plane-wave mode, and higher-order modes that vary in three-dimensions. These modes travel in both directions unless the duct termination is anechoic. Above a well-defined frequency, the contribution from higher-order modes can become significant. Experimental determination of plane-wave mode amplitudes using two microphones and broadband excitation was first proposed as an alternative to the impedance tube method for measuring acoustic impedance [3]. Mode decomposition of higher-order modes has been performed experimentally for a square-section duct which approximates the thin annulus of a high bypass-ratio engine [4]. For a circular-section duct, [5] included higher-order modes in their analysis of the sound produced by a fan.

Mode decomposition can be applied on both the inlet and outlet sides of a ducted device to determine the total amplitudes of waves travelling in each direction. However, some of the waves travelling away from the source may not be emanating directly from the device as reflection at, and transmission through, the device may occur. A source such as a fan within a duct with two openings or "ports" (i.e. inlet & outlet) can be modelled as a two-port source using a system of equations which relate its input and output states. The so-called "scattering matrix" formulation relates the pressure wave amplitudes on the inlet and outlet sides of the source and was first introduced by Davies [6] for plane-waves. Both the active (sound source) and passive (reflective/transmissive) properties of the device are systematically quantified and the method has been successfully used to characterise fans [7], mixer/orifice plates [8], [9] and other devices at relatively low frequencies. The inclusion of higher-order modes in these models was first done by Lavrentjev et al. [5] for a single port of a fan and has been used to study installation effects [10].

In this paper we present a two-port analysis that takes into account higher-order propagation modes to characterise an acoustic source. The mode decomposition method that is necessary for this analysis is presented in a new compact form amenable to optimisation of the locations of the (nonintrusive) measurements and for extension to include arbitrary numbers of propagating modes. The mode decomposition and source data from both sides of a test case fan are related using an extended scattering matrix which accounts for transmission and mode number changes. This enables the transmission of a spinning mode through a fan to be investigated to understand the observation that a mode spinning in the opposite sense to the rotor transmits less readily [11], [12]. Based on the measured source data for each mode the overall sound power is determined by integrating over the duct cross-section.

The measurements are compared to those taken in a rig built to the ISO standards [2], [13]. In the ISO standard rig, the sound power is calculated from the mean sound pressure from fixed radius inflow microphones on the inlet and outlet sides of the fan using a plane-wave formula. An analysis is performed to understand the accuracy obtained using measurements from a fixed radius at frequencies where non-planar modes can propagate.

In addition to potential benefits in terms of accuracy, the new method gives additional insight into the modal composition of the emitted sound. Anechoic terminations are not needed which can be large, expensive to build for a range of duct diameters and ineffective at some frequencies. By comparison the ISO standard rig is around 14 meters long while the experimental rig built for this study is only around 2.5 meters long. Locating the measurements on the duct surface is advantageous as it is nonintrusive and easy to define accurately.

## II. THEORY

### A. Mode Decomposition of Sound Emitted Into a Duct

Within a duct of circular cross-section of radius $a$ with no mean flow (for simplicity), the acoustic pressure $p$ satisfies the wave equation:

$$\frac{1}{c_0^2}\frac{\partial^2 p}{\partial t^2} - \nabla^2 p = 0 \tag{1}$$

By considering a solution with separable functions in cylindrical coordinates $(r, x, \theta)$ and applying appropriate boundary conditions, the acoustic pressure due to perturbations of frequency $\omega$ is:

$$p(r, x, \theta, t) = C_{m,n} e^{i(\omega t + m\theta)} J_m\left(z_{m,n}\frac{r}{a}\right)\left(A_{m,n}^+ e^{-ik_{m,n}x} + A_{m,n}^- e^{ik_{m,n}x}\right) \tag{2}$$

where $(m, n)$ are integers that represent azimuthal and radial mode number respectively, Bessel function of order $m$ are denoted $J_m$ and the zeros of its derivative $z_{m,n}$, and $A_{m,n}^+$ & $A_{m,n}^-$ are the wave amplitudes of mode $(m, n)$ travelling in the positive, and negative $x$ direction, respectively. The axial wave number $k_{m,n}$ is given by:

$$k_{m,n} = \sqrt{k_0^2 - \frac{z_{m,n}^2}{a^2}} = \sqrt{\frac{\omega^2}{c_0^2} - \frac{z_{m,n}^2}{a^2}} \tag{3}$$

The normalisation factor $C_{m,n}$ [14, Sec. 9.2] is set such that:

$$C_{m,n}^2 \oiint J_m^2\left(z_{m,n}\frac{r}{a}\right) ds = s \tag{4}$$

to make the amplitude representative of the mode power once the function is integrated over the cross-section (when finding the overall power – see Eq. (17)). This expression can be simplified (see appendix A) by performing the integration over a duct cross-section to give:

$$C_{m,n} = \left|\frac{z_{m,n}}{J_m(z_{m,n})\sqrt{(z_{m,n}^2-m^2)}}\right| \tag{5}$$

Assuming a uniform flow in the axial direction gives a similar result except that the axial wave number now depends on $M$, the mean-flow Mach number, and on the direction of wave propagation, either with (+) or against (-) the mean-flow

$$k_{m,n}^{\pm} = \frac{\sqrt{k_0^2-(1-M^2)\frac{z_{m,n}^2}{a^2}} \mp Mk_0}{1-M^2} \tag{6}$$

For frequencies below the cut-on frequency of a given mode, the axial wave number is purely imaginary and the amplitude of the wave decays exponentially along the duct axis. Conversely, above cut-on the wave can propagate without decay.

Taking the Fourier transform in time of (2) and summing over all possible modes gives:

$$\hat{p}(r,x,\theta,\omega) = \sum_{n=0}^{\infty}\sum_{m=-\infty}^{\infty} C_{m,n} e^{im\theta} J_m\left(z_{m,n}\frac{r}{a}\right)\left(A_{m,n}^+(\omega)e^{-ik_{m,n}x} + A_{m,n}^-(\omega)e^{ik_{m,n}x}\right) \tag{7}$$

In the present paper the three modes (0,0) & (±1,0) are measured experimentally. However, the technique presented here can be extended in a straightforward manner to additional modes.

From Eq. (7), the frequency-domain summation over the first three cut-on modes at a measurement location $(a, x_i, \theta_i)$ on the duct surface is given by:

$$\hat{p}_i(a, x_i, \theta_i, \omega) = \sum_{m=-1}^{m=1} C_{m,0} e^{im\theta_i} J_m(z_{m,0})\left(A_{m,0}^+(\omega)e^{-ik_{m,0}x_i} + A_{m,0}^-(\omega)e^{ik_{m,0}x_i}\right) \quad (8)$$

For each mode, there are two unknown amplitudes $A_{m,n}^+$ and $A_{m,n}^-$. This suggests that we require a minimum of two independent measurements, giving two equations, for each mode present. For the case presented here, there are two unknown amplitudes for each of $m = -1, 0, 1$ giving a minimum of six independent measurements ($N$ modes implies $2N$ unknowns). Denoting the six measurements required $\hat{p}_I, \hat{p}_{II}, \ldots, \hat{p}_{VI}$, the system of six simultaneous equations can be written in matrix form as:

$$\underbrace{\begin{bmatrix} \hat{p}_I(a, x_I, \theta_I, \omega) \\ \vdots \\ \hat{p}_{VI}(a, x_{VI}, \theta_{VI}, \omega) \end{bmatrix}}_{p} = \underbrace{\begin{bmatrix} C_{0,0}e^{-ik_{0,0}x_I} & \cdots & C_{-1,0}e^{i\theta_I} J_{-1}(z_{-1,0})e^{ik_{-1,0}x_I} \\ \vdots & \ddots & \vdots \\ C_{0,0}e^{-ik_{0,0}x_{VI}} & \cdots & C_{-1,0}e^{i\theta_{VI}} J_{-1}(z_{-1,0})e^{ik_{-1,0}x_{VI}} \end{bmatrix}}_{M} \underbrace{\begin{bmatrix} A_{0,0}^+(\omega) \\ \vdots \\ A_{-1,0}^-(\omega) \end{bmatrix}}_{A^{\pm}} \quad (9)$$

This set of simultaneous equations is solved at each frequency $\omega$ for the unknown amplitudes:

$$\{A^{\pm}\} = M^{-1}\{p\} \quad (10)$$

The matrix $M$ is singular or ill-conditioned if the six measurements are not fully independent. Computational simulations were carried out to determine the optimum measurement locations and sensitivity to errors. This is detailed later. Surplus measurements can be used to form additional entries in the $p$ vector and matrix $M$ to give an overdetermined system of equations to improve mode decomposition results [8]. Note that Eq. (10) works only for deterministic signals.

For a signal with noise, we should use average statistical quantities. The formulation was recast in terms of quantities to be averaged over a significant measurement period, valid for both periodic and random signals. Dividing Eq. (9) by a coherent reference pressure (or voltage) signal $\hat{r}$ and using the inverse of $\boldsymbol{M}$:

$$\begin{bmatrix} \frac{A_{0,0}^+}{\hat{r}} \\ \vdots \\ \frac{A_{-1,0}^-}{\hat{r}} \end{bmatrix} = \boldsymbol{M}^{-1} \begin{bmatrix} \widehat{H}_{rI} \\ \vdots \\ \widehat{H}_{rVI} \end{bmatrix} \tag{11}$$

where for $i = I, II, \ldots, VI$

$$\frac{\hat{p}_i}{\hat{r}} = \frac{\overline{\hat{p}_i \hat{r}^*}}{\overline{\hat{p}_r \hat{r}^*}} = \frac{\hat{S}_{ir}}{\hat{S}_{rr}} = \widehat{H}_{ri} \tag{12}$$

in which $\hat{S}_{ir}$ and $\hat{S}_{rr}$ represent cross- and auto-spectral density, respectively. This approach works only when the coherence is good between the pairs of signals [15].

The quantities of interest, such as the magnitudes of the mode amplitudes can then be found using the auto spectral density of the reference signal, for example:

$$\hat{S}_{A_{0,0}^+ A_{0,0}^+} = |A_{0,0}^+|^2 = \left|\frac{A_{0,0}^+}{\hat{r}}\right|^2 \hat{S}_{rr} \tag{13}$$

or the predicted auto spectral density anywhere in the duct using Eq. (8):

$$\hat{S}_{ii}(r_i, x_i, \theta_i, \omega) = |\hat{p}_i|^2 = \left|\frac{\hat{p}_i}{\hat{r}}\right|^2 \hat{S}_{rr} \tag{14}$$

Calculation of the net sound power travelling in a duct, taking into account higher-order modes, requires integration of the axial intensity field over a cross-section. The time-averaged axial intensity as a function of the acoustic pressure $\hat{p}$ (from (7)) and axial acoustic velocity $\hat{v}_x$ is given by:

$$\bar{I}_x(r, x, \theta, \omega) = \frac{1}{2} Re(\hat{p}\hat{v}_x^*) \tag{15}$$

This is then integrated over the cross-section area $s$ to find the sound power:

$$P = \oiint \bar{I}_x ds \tag{16}$$

It is shown in appendix B that the cross terms in the integration involving products of the amplitudes of different modes do not contribute to the sound power. Equation (16) can therefore be simplified for a given mode as:

$$P_{m,n} = \frac{sk_{m,n}}{2k_0\rho_0 c_0}(|A_{m.n}^+|^2 - |A_{m.n}^-|^2) \tag{17}$$

where the overall sound power is found by adding the contribution from each mode i.e. $P = \sum_{n=0}^{\infty}\sum_{m=-\infty}^{\infty} P_{m,n}$.

**B. Aeroacoustic source characterisation**

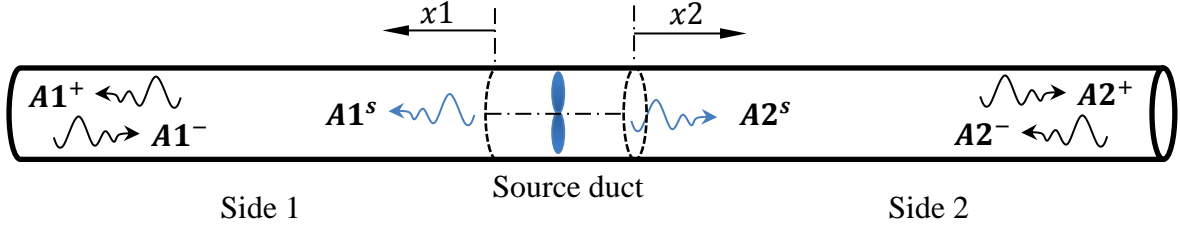

FIG. 1. (Color online) Two-port source wave amplitudes

A source such as a fan within a duct with two openings or "ports" can be modelled as a two-port source using a system of equations which relate its input and output states. This situation is illustrated in Fig. 1 for a given mode. The so-called "scattering matrix" formulation relates the pressure wave amplitudes on the inlet and outlet sides of the source and was first introduced by [6]. This allows the waves due to the source alone (denoted by superscript 's' in the figure) to be discerned from the total wave amplitudes travelling each way on either side of the source (denoted $A1^{\pm}$ & $A2^{\pm}$). At lower frequencies, only a single plane-wave mode can propagate on each side of the source which results in 2 simultaneous equations:

$$\underbrace{\begin{bmatrix} A1^+_{0,0} \\ A2^+_{0,0} \end{bmatrix}}_{A^+} = \underbrace{\begin{bmatrix} \rho 1 & \tau 21 \\ \tau 12 & \rho 2 \end{bmatrix}}_{S} \underbrace{\begin{bmatrix} A1^-_{0,0} \\ A2^-_{0,0} \end{bmatrix}}_{A^-} + \underbrace{\begin{bmatrix} A1^s_{0,0} \\ A2^s_{0,0} \end{bmatrix}}_{A^s} \qquad (18)$$

where the $S$ is the scattering matrix, $A^s$ is a source vector and the sign convention for the wave directions of travel is shown in Fig. 1. The scattering matrix contains four parameters characterising the reflective ($\rho$) and transmissive ($\tau$) properties of the source duct element. Note that the wave amplitudes are defined as illustrated in Fig. 1 for the specific case of the (0,0) mode. The extension to the first three modes gives 6 simultaneous equations and 36 unknowns in the scattering matrix:

$$\underbrace{\begin{bmatrix} A1_{0,0}^+ \\ A1_{1,0}^+ \\ A1_{-1,0}^+ \\ A2_{0,0}^+ \\ A2_{1,0}^+ \\ A2_{-1,0}^+ \end{bmatrix}}_{A^+} = \underbrace{\begin{bmatrix} \rho 1_{0,0 \to 0,0} & \cdots & \tau 21_{-1,0 \to 0,0} \\ \vdots & \ddots & \vdots \\ \tau 12_{0,0 \to -1,0} & \cdots & \rho 2_{-1,0 \to -1,0} \end{bmatrix}}_{S} \underbrace{\begin{bmatrix} A1_{0,0}^- \\ A1_{1,0}^- \\ A1_{-1,0}^- \\ A2_{0,0}^- \\ A2_{1,0}^- \\ A2_{-1,0}^- \end{bmatrix}}_{A^-} + \underbrace{\begin{bmatrix} A1_{0,0}^s \\ A1_{1,0}^s \\ A1_{-1,0}^s \\ A2_{0,0}^s \\ A2_{1,0}^s \\ A2_{-1,0}^s \end{bmatrix}}_{A^s} \qquad (19)$$

where the terms have a degree of symmetry about the leading diagonal coefficients (corresponding to a proportion of reflection without a change in the mode number). For example, the coefficient $\tau 12_{0,0 \to -1,0}$ corresponds to transmission of a wave from side 1 to 2 in which the mode number changes (from 0 to -1). Dividing through by a coherent reference signal $\hat{r}$ gives a form where the amplitudes follow from mode decomposition on sides 1 and 2 using Eq. (11):

$$\begin{bmatrix} \frac{A1_{0,0}^+}{\hat{r}} \\ \vdots \\ \frac{A2_{-1,0}^+}{\hat{r}} \end{bmatrix} = S \begin{bmatrix} \frac{A1_{0,0}^-}{\hat{r}} \\ \vdots \\ \frac{A2_{-1,0}^-}{\hat{r}} \end{bmatrix} + \begin{bmatrix} \frac{A1_{0,0}^s}{\hat{r}} \\ \vdots \\ \frac{A2_{-1,0}^s}{\hat{r}} \end{bmatrix} \qquad (20)$$

Note that as in Eq. (11) the vector components of the above represent averaged statistical quantities. Using an external source such as a loudspeaker, the scattering matrix $S$ is found experimentally by producing sound fields in the duct of much higher amplitude and uncorrelated with the source vector $A^s$ in (18), which becomes effectively zero.

When including the first three modes for which there are six unknown modal amplitudes contained in the vectors $A^\pm$ (as in Eq. (9)), the scattering matrix $S$ contains thirty six parameters in a six-by-six matrix. At least six independent sound fields are required to solve for the matrix $\mathbf{S}$ (i.e. $2N$ sound fields for $N$ modes). The solution procedure is discussed in more detail in §III.

Once $\mathbf{S}$ is known, the external sources are deactivated and the source vector of interest can be found. The source sound power from a mode, based on the source vector $\mathbf{A}^s$, follows from Eq. (17):

$$P_{m.n}^s = \frac{sk_{m,n}}{2k_0\rho_0 c_0}|A_{m.n}^s|^2 \tag{21}$$

and the overall source sound power is found by summing the contribution from each mode i.e. $P^s = \sum_{n=0}^{\infty}\sum_{m=-\infty}^{\infty} P_{m.n}^s$. For the case of plane-waves only for which $k_{m,n} = k_0$ Eq. (21) reduces to:

$$P^s = \frac{s}{2\rho_0 c_0}|A^s|^2 \tag{22}$$

which can be rearranged into a formula relating sound power to sound pressure level measured directly under anechoic conditions (as per §8.2 in the ISO standard [2]):

$$PWL = 10\log\left(\frac{P^s}{PWL_{ref}}\right) = \underbrace{20\log\frac{|A^s|_{rms}}{SPL_{ref}}}_{SPL} + 10\log s - 10\log\left(\frac{\rho_0 c_0}{(\rho c)_{ref}}\right) \tag{23}$$

where $PWL_{ref}$ is 1 pW, $SPL_{ref}$ is 20 µPa and $(\rho c)_{ref} = SPL_{ref}^2/PWL_{ref}$.

## III. METHODOLOGY

### A. Measurement location optimisation

In order to optimise the locations of the six pressure measurements over a wide range, an optimisation procedure was developed in MATLAB. This tested every possible combination of azimuthal angles with six measurement locations equally split between planes at two axial locations $x_A, x_B$ (for ease of construction). The aim of the optimisation is to find the azimuthal configuration which minimises the condition number $\kappa[\mathbf{M}]$ at the high end of the frequency range of interest for the present work (2 kHz). The condition number quantifies how independent the six measurements are; a lower condition number implies a higher independence of the measurement set. During the first phase of the optimisation, the axial spacing between the planes was fixed i.e.

$$x_i = \begin{cases} x_A, & i = I - III \\ x_B, & i = IV - VI \end{cases} \tag{24}$$

In this case the optimisation process for the azimuthal locations can be represented mathematically as:

$$\min_{\theta_i \in [0, 2\pi]} \kappa[\mathbf{M}(x_B - x_A, \theta_i, \omega)] \tag{25}$$

assuming that the ideal axial spacing would be close to the theoretical optimum for the plane-wave only case corresponding to a quarter of the wavelength [16] i.e. $x_B - x_A = \lambda_{2kHz}/4$. Afterwards, this assumption was relaxed by fixing the ideal azimuthal distribution and then varying the axial spacing between zero and half wavelength spacing:

$$\min_{x_B - x_A \in [0, \frac{\lambda_{2kHz}}{2}]} \kappa[\boldsymbol{M}(x_B - x_A, \theta_i, \omega)] \qquad (26)$$

The ideal azimuthal configuration was found to be equal angular spacing. However the optimum axial spacing was found to be higher than the quarter wavelength ideal for the plane-wave region (see Fig. 2 (a)). The condition number for this ideal arrangement at lower frequencies is shown in Fig. 2 (b).

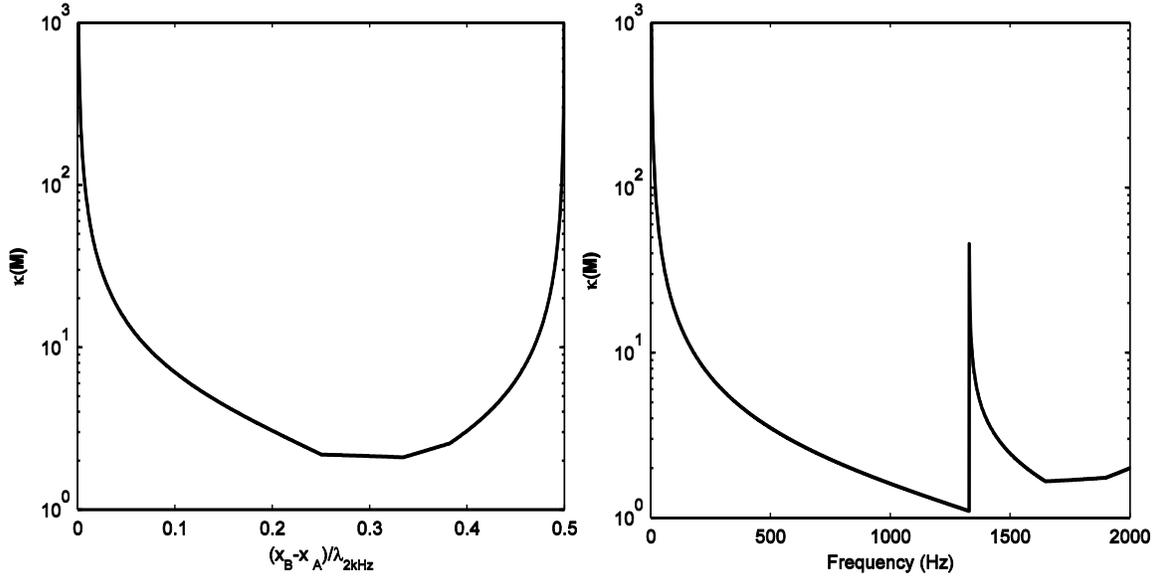

FIG. 2. (a) Variation of condition number $\kappa$ of modal matrix $\boldsymbol{M}$ with axial spacing normalised by wavelength at 2 kHz, minimum indicated (b) Variation of condition number $\kappa$ for the ideal axial and azimuthal arrangement at the range of frequencies of interest

The peak in condition number visible in Fig. 2 (b) around 1.3 kHz corresponds to the cut-on frequency for the first azimuthal mode and is due to the axial wave number being close to zero. However, this does not affect the results significantly (see Fig. 7).

**B. Aeroacoustic source characterisation rig set-up**

For both types of tests, with a loudspeaker source and with a fan, the basic set-up of the experimental rig was the same. A schematic of the rig layout is shown in Fig. 3 and the actual rig is pictured in Fig. 4. The rig is significantly shorter (c. 2.5 m) than the ISO standard rig of the same diameter used later (c. 14 m).

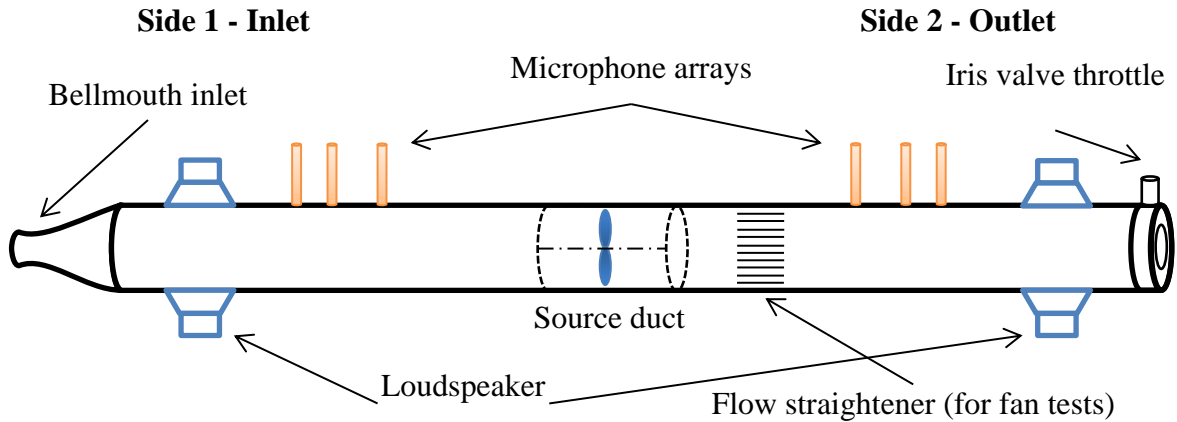

FIG. 3. (Color online) Schematic of measurement rig layout

A loudspeaker source driven by broadband noise was located in the source duct element for the initial tests to validate the method in the absence of flow. To determine the flow rate through the rig during fan tests, an ISO standard bellmouth inlet was constructed [13] and integrated upstream. The operating point of the fan was accurately controlled using an iris valve and stepper motor at the outlet to throttle the flow, similar to an orifice plate of variable area.

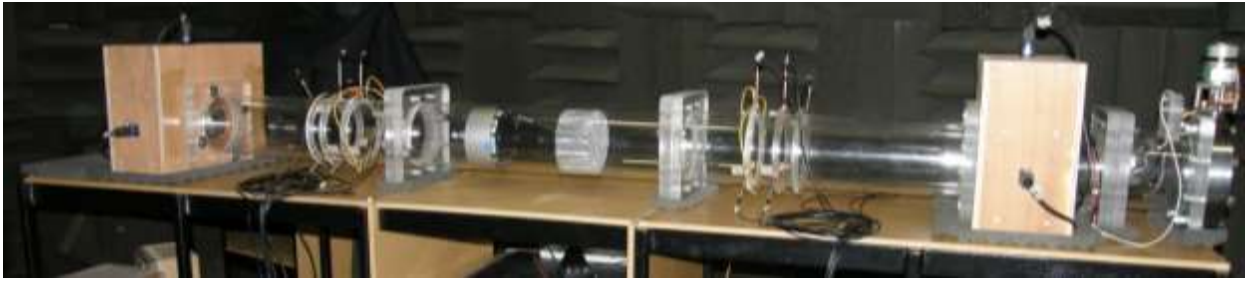

FIG. 4. (Color online) Picture of measurement rig layout with test case fan and flow straightener installed

Fig. 5 shows the non-dimensional operating curves of the fan measured in both rigs, and the design point at which it operated during the study. The fan was driven directly by an electric motor and controller which kept the rotational speed constant with high accuracy. The Mach number of the mean velocity in the duct was very low (~ 0.01) so that the acoustic theory for no mean flow was applicable.

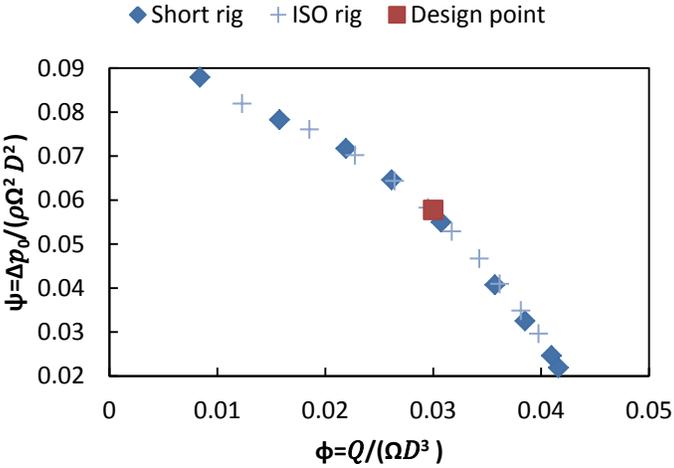

FIG. 5. (Color online) Non-dimensional pressure rise $\psi$ of test case fan as a function of flow coefficient $\phi$ with comparison to the performance measured in the ISO rig

The microphones chosen for surface pressure measurements were G.R.A.S. eighth-inch condenser microphones (Type 40DD) for high spatial resolution. Data was acquired from the 14 microphones and reference signals using a GBM Viper multichannel system capable of simultaneous data acquisition, signal conditioning for optimal ADC resolution and anti-aliasing filtering, and sensor power supply.

**C. Flow uniformity and noise reduction**

The unsteady turbulent pressure fluctuations at the measurement locations must be lower than the acoustic pressure fluctuations for accurate source characterisation at the frequencies of interest. The ISO standard method [2] requires that this difference should be greater than 6 dB and specifies a range of test procedures (e.g. inflow microphone shielding) which depend on the mean velocity and swirl component. An advantage of the theory detailed in §II is that it is based on cross-correlations between microphone signals and a reference signal. In some parts this reference is a noise-free loudspeaker driving signal. In other cases the reference is another flush-mounted microphone with which localised turbulence effects should not be well correlated.

The accuracy of the mode decomposition predictions was monitored by using Eq. (14) to predict the (cross) spectra measurement at an arbitrary location in the duct. This prediction was then compared to the actual measurement made simultaneously at that location to highlight any discrepancies in terms of magnitude (or phase). Frequencies at which there was low coherence between measurements were also discarded.

A honeycomb flow straightener was used downstream of the fan to remove the swirl component of the mean flow which can cause excess flow noise. As discussed in the ISO standard [2], a straightener can potentially generate excess noise hence it is recommended that readings are taken with and without it, and the minimum is used. In order to quantify the effect of the straightener on the sound travelling towards the measurement locations, tests were performed to measure the passive properties of the flow straightener alone.

### D. Scattering matrix determination

For the two-port analysis, the scattering matrix $S$ was found using external excitation from arrays of loudspeakers mounted on the surface of the duct near the inlet and outlet of the rig. The duct surface was perforated to enable sound to be emitted into the duct without disturbing the flow. By varying the location and phases of the active speakers, several independent sound fields could be generated in the duct. The uncorrelated parts of the measured pressure signals were suppressed using the loudspeaker driving voltage as the reference signal in (11) and averaging over many spectra using Welch's method [17]. The use of single frequency sine wave excitation gave a high coherence with all microphones and degree of control. This is particularly important when measuring a source with very low transmission for which it is difficult to achieve high coherence between the loudspeaker signal and a microphone measuring on the opposite side of the source. The excitation can be automatically stepped in 5 Hz increments to cover a range of frequencies.

Fig. 6 shows the 90° azimuthal spacing of the array speakers. To preferentially excite the first azimuthal mode, the speakers were excited with a 90° phase shift between the driving signals as summarised in Table I. Six sound fields were generated by exciting the inlet and outlet arrays in turn.

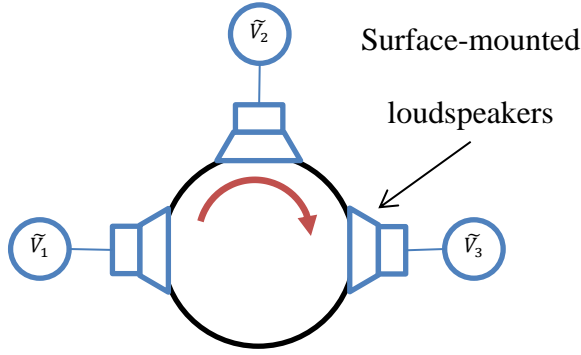

FIG. 6. (Color online) Illustration of the azimuthal arrangement of the speakers where the speakers are driven with a phase shift between each to (preferentially) excite an azimuthal mode

TABLE I. Independent phase settings of the speaker excitation voltages to preferentially excite different modes

|             | Relative phase of speaker |             |             |
| ----------- | ------------ | ----------- | ----------- |
| Combination | $\tilde{V}_1$ | $\tilde{V}_2$ | $\tilde{V}_3$ |
| 1           | 0            | +90         | +180        |
| 2           | 0            | -90         | -180        |
| 3           | 0            | 0           | 0           |

The sets of data from each excitation set form a vector in the matrices of mode amplitudes (from Eq. (20) with $\boldsymbol{A^s}/\hat{r} = 0$):

$$\begin{bmatrix} \uparrow & & \uparrow \\ \frac{\boldsymbol{A^+}}{\hat{r}}_I & \cdots & \frac{\boldsymbol{A^+}}{\hat{r}}_{VI} \\ \downarrow & & \downarrow \end{bmatrix} = \boldsymbol{S} \begin{bmatrix} \uparrow & & \uparrow \\ \frac{\boldsymbol{A^-}}{\hat{r}}_I & \cdots & \frac{\boldsymbol{A^-}}{\hat{r}}_{VI} \\ \downarrow & & \downarrow \end{bmatrix} \qquad (27)$$

from which **S** can be found by inverting $A^-$. It is difficult to excite modes independently using he speaker installation described above. This is because of several reasons: the sound field from the speakers enter the duct through perforations. This yields a distributed source and so we do not have excitation at discrete point sources that are 90° out of phase required for excitation of a pure spinning mode. Also, for simplicity, we don't have a fourth speaker at the bottom needed to complete the symmetry of the excitation. This resulted in some level of dependence between sets. This problem was solved by using an additional speaker at a different axial location to give another data set. Equation (27) was then solved as an overdetermined system.

## IV.  RESULTS

### A. Mode decomposition

Fig. 7 shows the comparison of prediction and measured spectra at the verification location. The agreement is good apart from at amplitudes below the lower end of the dynamic range of the microphones (40 dB) or when assuming only plane-waves.

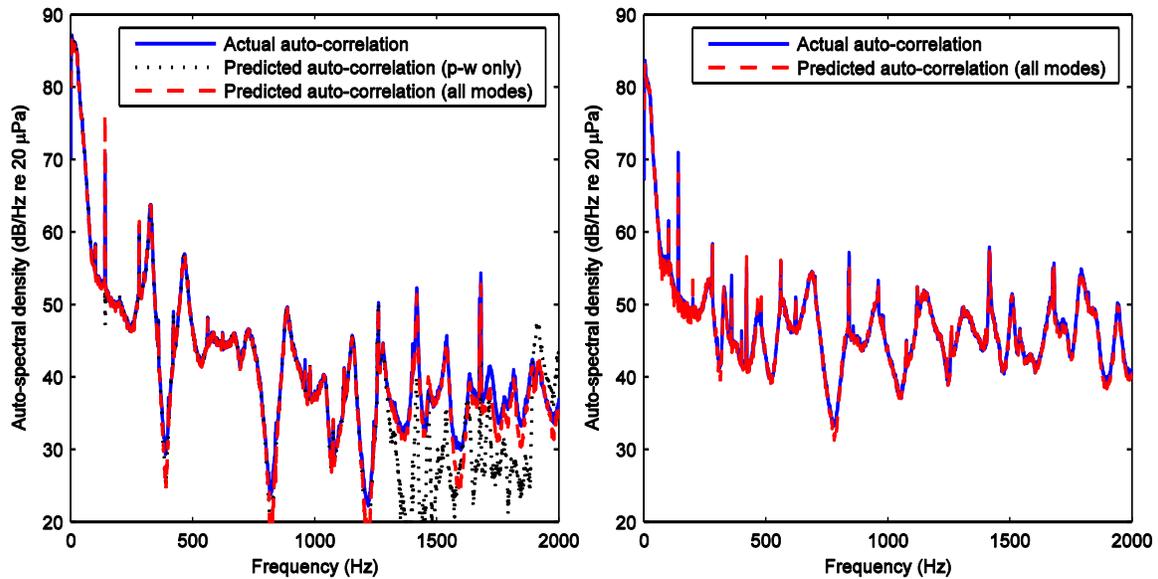

FIG. 7. (Color online) Comparison between predictions and actual measurements for fan at the verification location on the (a) inlet side 1 (b) outlet side 2

The random error in the cross spectral estimates used for mode decomposition can be shown [15] to depend on $\left(\frac{1}{2n}\right)\left[\left(\frac{1}{\gamma^2}\right) - 1\right]$ where $n$ is the number of spectral averages and $\gamma^2$ is the magnitude squared coherence. The number of spectra averages was 256 and coherence values for the tone of interest are shown in Table II. As a guide, the ISO standard states that the turbulent pressure fluctuations should be at least 6 dB lower than the acoustic fluctuations in which case the coherence must be greater than 0.64 (for plane-wave frequencies) [2].

TABLE II. Magnitude squared coherence values with 256 spectra averages

|  | Average coherence estimate with reference signal across all 6 measurements | |
|---|---|---|
| Frequency | Side 1 | Side 2 |
| 1.68 kHz | 0.94 | 0.86 |

**B. Source characterisation of mixed-flow fan**

We had recently characterised the sound from a mixed-flow fan using only two external sources necessary for the plane-wave frequencies [18]. Each external source, located on the inlet and outlet sides, was driven in turn using broadband noise to cover a range of frequencies simultaneously. The magnitudes of the reflection and transmission coefficients in the scattering matrix are shown in Fig. 8. In the absence of flow, the reciprocity principle implies that the transmission coefficients should be equal. Due to the low Mach number of the flow, it is evident from Fig. 8 that there are only small deviations from this principle.

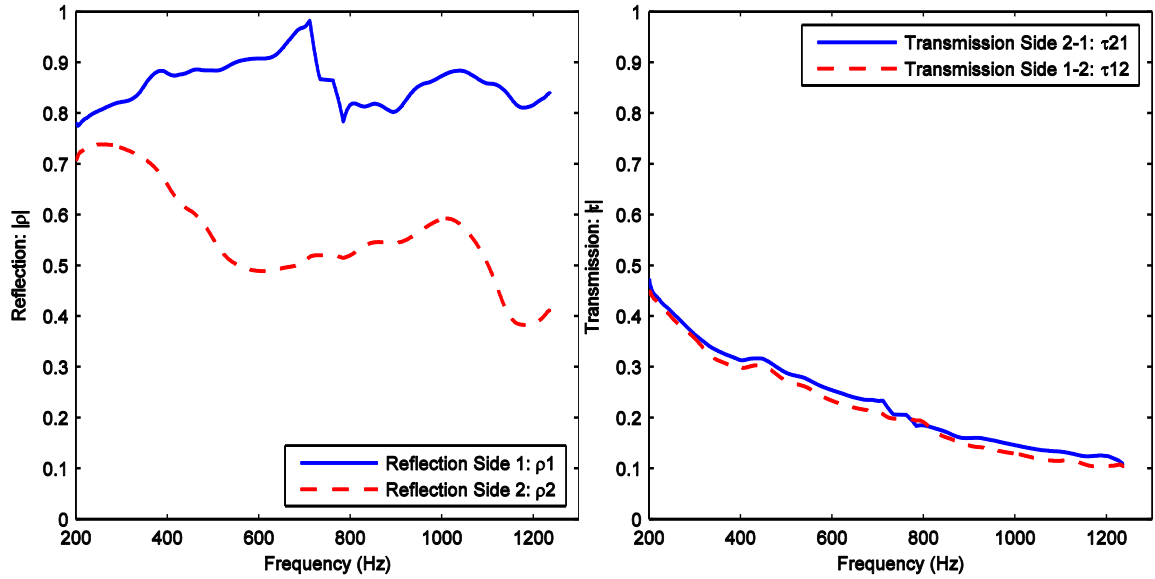

FIG. 8. (Color online) Scattering matrix data at plane-wave frequencies for test case fan operating at its design point: (a) reflection (b) transmission

In this work, the method has been extended and used to characterise the first significant tone at 1.68 kHz (see Fig. 7, most other peaks above 1.3 kHz are due the repeating resonance pattern) which occurs in a range of frequencies for which higher-order modes propagate in the duct. As discussed in §III, this requires the generation of six independent sound fields. It is also clear from Fig. 8 that the general trend is for transmission to reduce at higher frequency. For these reasons, excitation at a single frequency was required for controllability and high signal-to-noise ratios (particularly for measurements on the opposite side to the external excitation).

The results from using the external sources to characterise a mixed-flow fan operating at its design point are presented below. The magnitudes of the reflection and transmission coefficients in the scattering matrix are given in Table III. It is clear that in general the transmission through the fan at 1.68 kHz is low while reflection is relatively high, particularly at the inlet (side 1). In each of the four quadrants of the table the largest values are seen on the leading diagonal which corresponds to reflection/transmission without mode number change.

TABLE III. Scattering matrix $S$ data at 1.68 kHz for test case fan operating at its design point

|  | Reflected as | | | Transmitted as | | |
| --- | --- | --- | --- | --- | --- | --- |
|  | (0,0) | (1,0) | (-1,0) | (0,0) | (1,0) | (-1,0) |
| Incoming (0,0) to Side 1 | 0.75 | 0.20 | 0.36 | 0.03 | 0.04 | 0.01 |
| Incoming (1,0) to Side 1 | 0.12 | 0.84 | 0.19 | 0.01 | 0.06 | 0.01 |
| Incoming (-1,0) to Side 1 | 0.10 | 0.04 | 0.91 | 0.02 | 0.00 | 0.05 |
| Incoming (0,0) to Side 2 | 0.69 | 0.08 | 0.14 | 0.05 | 0.02 | 0.01 |
| Incoming (1,0) to Side 2 | 0.06 | 0.57 | 0.04 | 0.01 | 0.04 | 0.02 |
| Incoming (-1,0) to Side 2 | 0.04 | 0.13 | 0.56 | 0.01 | 0.01 | 0.05 |

Based on the scattering matrix data the amplitudes of the source $A^s$ were calculated by deactivating the external sources and measuring the sound field produced by the fan. The source data are presented in Table IV.

TABLE IV. Source data for each mode at 1.68kHz produced by test case fan

|  | Mode amplitude (dB/Hz re. 20μPa) | |
|---|---|---|
|  | Side 1 | Side 2 |
| Plane-wave mode (0,0) | 42.68 | 55.11 |
| Azimuthal mode (1,0) | 38.47 | 45.72 |
| Azimuthal mode (-1,0) | 40.83 | 47.25 |

**C. Influence of honeycomb flow straightener**

The acoustic effect of the honeycomb flow straightener was investigated by characterising its passive properties (i.e. the scattering matrix) in isolation by removing the fan from the source duct element. Ideally this should have very low reflection and high transmission. Table V shows the scattering matrix data.

TABLE V. Scattering matrix data for the honeycomb flow straightener at 1.68 kHz

|  | Reflected as | | | Transmitted as | | |
|---|---|---|---|---|---|---|
|  | (0,0) | (1,0) | (-1,0) | (0,0) | (1,0) | (-1,0) |
| Incoming (0,0) to Side 1 | 0.12 | 0.03 | 0.23 | 0.90 | 0.18 | 0.06 |
| Incoming (1,0) to Side 1 | 0.03 | 0.30 | 0.04 | 0.05 | 0.84 | 0.02 |
| Incoming (-1,0) to Side 1 | 0.05 | 0.01 | 0.33 | 0.03 | 0.03 | 0.79 |
| Incoming (0,0) to Side 2 | 0.06 | 0.11 | 0.02 | 0.95 | 0.02 | 0.11 |
| Incoming (1,0) to Side 2 | 0.06 | 0.40 | 0.05 | 0.03 | 0.83 | 0.03 |
| Incoming (-1,0) to Side 2 | 0.01 | 0.02 | 0.36 | 0.04 | 0.05 | 0.80 |

By considering the sum of the square of the magnitudes of the reflection coefficients, it was clear that only a very small proportion of the acoustic power is reflected by the honeycomb. For the plane-wave mode which contributes most to the power, at most 7% is reflected and this would only change the power prediction for this mode on the order of 0.3 dB. For the azimuthal modes, at most 17% is reflected and this would have an effect on the order of 0.8 dB for these modes. Consequently, the outlet side sound power was expected to be accurate to around ± 1 dB due to the presence of the honeycomb flow straightener.

**D. Comparison with ISO rig measurements of mixed-flow fan**

The sound power was determined using Eq. (21). The power levels from the ISO rig were calculated as indicated in the ISO standard using the plane-wave formula to relate sound pressure and sound power (see Eq. (23)). Table VI and Table VII show a comparison of the results. The effect of the flow straightener in the ISO rig (of the "eight-radial-vane" design [13]) on the outlet sound power prediction was not quantified. The right-most column in these tables represent the SPL that was used with the plane-wave formula to find the PWL in the ISO rig.

TABLE VI. Comparison between the sound power level estimations at 1.68 kHz made with the ISO rig – fan inlet side 1

| **PWL New rig** (dB/Hz re. 1pW) | **PWL ISO rig** (dB/Hz re. 1 pW) | **Equivalent ISO rig SPL** (dB/Hz re. 2 μPa) |
|---|---|---|
| 23.6 | 23.5 | 41.2 |

TABLE VII. Comparison between the sound power level estimations at 1.68 kHz made with the ISO rig – fan outlet side 2

| **PWL** (dB/Hz re. 1pW) | **ISO rig PWL** (dB/Hz re. 1 pW) | **Equivalent ISO rig SPL** (dB/Hz re. 2 µPa) |
|---|---|---|
| 34.7 | 30.8 | 48.5 |

The potential inaccuracies from taking a single measurement at $r/a = 0.5$ to represent the sound pressure level can be understood by looking at Fig. 9 and Fig. 10 which show the spatial variation of the predicted SPL at an axial location on the inlet and outlet sides, respectively. The relationship between sound pressure and power used in the ISO standard are derived in Eq. (23) and the equivalent sound pressure levels used to calculate the sound power level for the ISO rig are included in Table VI and Table VII. Considering the measurements made at $r/a = 0.5$ and three equally-spaced angles of 90°,210°,330° on the inlet side, the SPL measurements under anechoic conditions would be 48.2, 42.4 and 28.4 dB/Hz respectively. The magnitude-squared average of these is 44.5 dB/Hz which is several decibels higher than the equivalent SPL level in Table VI. This sort of variation is to be expected since it is clear from Fig. 9 that the measurements depend strongly on the absolute azimuthal location.

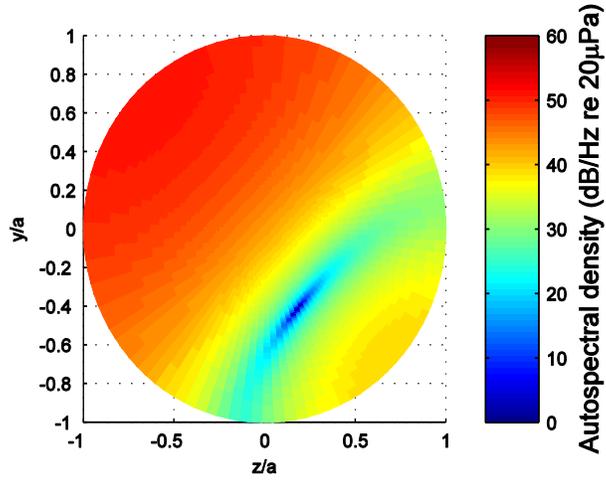

FIG. 9. (Color online) Variation of the predicted SPL (autospectral density) that would be measured under anechoic conditions for a duct cross-section on the inlet side

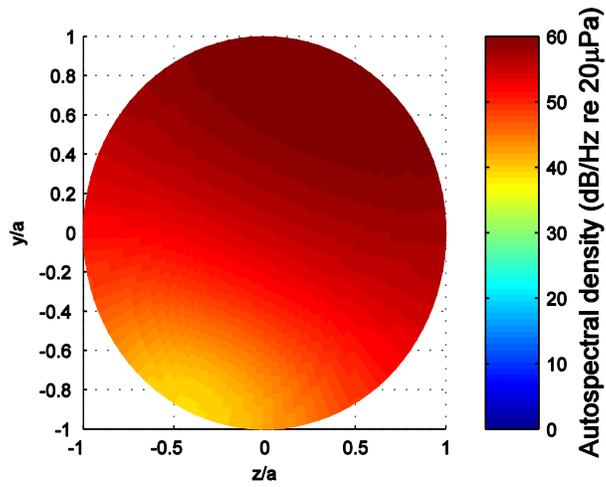

FIG. 10. (Color online) Variation of the predicted SPL (autospectral density) that would be measured under anechoic conditions for a duct cross-section on the outlet side

## V. CONCLUSIONS

A technique has been developed and validated to decompose higher-order modes of sound emitted into a duct. Due to the increase in independent measurements required, a computational method was used to simulate sensitivities of real measurements and optimise the set-up. The 'two-port' source model for a source with two active openings has been formulated to include higher-order modes and applied for the first three modes as a proof of concept. This involved constructing an experimental rig with six independent measurements from surface, flush-mounted microphones on the inlet and outlet sides of the source. The rig was designed to enable aerodynamic and acoustic performance measurements of small ducted fans in a similar way to rigs built to the ISO standard [2]. A particular advantage of the proposed methodology is that the rig is much shorter than the ISO rig and does not require anechoic terminations. A fan outlet flow straightener has been characterised to understand its effect on subsequent acoustic measurements.

Comparison with measurements of narrow-band sound power made on an ISO rig indicates that any ISO rig may not yield accurate results when higher-order modes propagate. Investigations of the underlying structure of the sound field with higher-order modes suggest that measurements of sound pressure level at a fixed radius (between the centre and walls) may not accurately represent the field which can vary strongly with radius and azimuth. Another source of inaccuracy in an ISO rig is that they may not be fully anechoic. We found this to be the case with the ISO rig we had used in an earlier study (see [18]). As the source power calculation in an ISO rig is based on the assumption that there are no reflections from the end of the duct, these reflections are another source of error.

## VI. ACKNOWLEDGEMENTS

The financial and technical support, and collaboration from Dyson Ltd. has made this research possible, and we are particularly thankful to Ryan Stimpson of the aeroacoustic RDD team. The technical support of John Hazelwood at Cambridge is also greatly appreciated.

## VII. APPENDICES

### A. Simplification of Bessel function integral

Integrating over a circular cross-section $s$ with elemental area $ds = 2\pi r dr$ and radius $a$

$$\oiint J_m^2\left(z_{m,n}\frac{r}{a}\right) ds = 2\pi \int_0^a J_m^2\left(z_{m,n}\frac{r}{a}\right) r dr \tag{28}$$

Using the ascending series definition for $J_m$ and the recurrence relation relating $J_m$ and $J_{m-1}$ (see [19] for example) this integral can be expressed in terms of Bessel functions

$$2\pi \int_0^a J_m^2\left(z_{m,n}\frac{r}{a}\right) r dr = s\frac{J_{m-1}^2(z_{m,n}) - 2mJ_{m-1}(z_{m,n})J_m(z_{m,n}) + zJ_m^2(z_{m,n})}{z_{m,n}} \tag{29}$$

From the definition for the zeroes of $J_m'(z)$ denoted $z_{m,n}$ and the recurrence relation for this differential:

$$J_m'(z_{m,n}) = J_{m-1}(z_{m,n}) - \frac{mJ_m(z_{m,n})}{z_{m,n}} = 0 \tag{30}$$

Substituting for $J_{m-1}$ in the right-hand side of Eq. (29) using this result simplifies the integral:

$$\oiint J_m^2\left(z_{m,n}\frac{r}{a}\right) ds = sJ_m^2(z_{m,n})\frac{z_{m,n}^2 - m^2}{z_{m,n}^2} \tag{31}$$

The definition for the normalisation factor, Eq. (4), can be written as

$$C_{m,n}^2 \oiint J_m^2\left(z_{m,n}\frac{r}{a}\right)ds = s \tag{32}$$

which leads to the final expression for $C_{m,n}$ by substituting the simplified expression

$$C_{m,n} = \left|\frac{z_{m,n}}{J_m(z_{m,n})\sqrt{(z_{m,n}^2-m^2)}}\right| \tag{33}$$

**B. Sound power integral**

The acoustic pressure and velocity of the sound field are linked by the linear conservation of momentum equation. The axial component of this equation leads to the following relations for waves travelling in the positive $x$ direction only:

$$\hat{v}_x = \frac{i}{\omega\rho_0}\frac{\partial\hat{p}}{\partial x} = \frac{k_{m,n}}{k_0 c_0 \rho_0}\hat{p} \tag{34}$$

and from Eq. (7)

$$\hat{p} = \sum_{n=0}^{\infty}\sum_{m=-\infty}^{\infty} C_{m,n} e^{im\theta} J_m\left(z_{m,n}\frac{r}{a}\right) A_{m,n}^+(\omega) e^{-ik_{m,n}x} \tag{35}$$

The time-averaged axial intensity is therefore, assuming the modes are cut-on (i.e. $k_{m,n}$ is real), given by:

$$\bar{I}_x = \frac{1}{2}Re(\hat{p}\hat{v}_x^*) = \frac{1}{2}Re\left(\frac{k_{m,n}}{k_0 c_0 \rho_0}\hat{p}\hat{p}^*\right) = \sum_{n=0}^{\infty}\sum_{m=-\infty}^{\infty}\frac{k_{m,n}}{2k_0 c_0 \rho_0}C_{m,n}^{\;2}J_m^{\;2}\left(z_{m,n}\frac{r}{a}\right)|A_{m.n}^+|^2 \tag{36}$$

The power in one direction is found by integrating over a duct cross-section:

$$P^+ = \oiint \bar{I}_x ds = \sum_{n=0}^{\infty}\sum_{m=-\infty}^{\infty}\frac{k_{m,n}}{2k_0 c_0 \rho_0}|A_{m.n}^+|^2\left[C_{m,n}^{\;2}\oiint J_m^{\;2}\left(z_{m,n}\frac{r}{a}\right)ds\right] =$$

$$\sum_{n=0}^{\infty}\sum_{m=-\infty}^{\infty}\frac{sk_{m,n}}{2k_0 c_0 \rho_0}|A_{m.n}^+|^2 \tag{37}$$

where the term in square brackets simplifies to simply $s$ from Eq. (32). The net sound power is found by including the waves travelling in the negative $x$ direction

$$P = \sum_{n=0}^{\infty}\sum_{m=-\infty}^{\infty}\frac{sk_{m,n}}{2k_0 c_0 \rho_0}(|A_{m.n}^+|^2 - |A_{m.n}^-|^2) \tag{38}$$

As expected the total power is a summation of the power contribution from each mode $P_{m,n}$, which can be written:

$$P_{m,n} = \frac{sk_{m,n}}{2k_0 c_0 \rho_0}(|A_{m.n}^+|^2 - |A_{m.n}^-|^2) \tag{39}$$

FIG. 1. (Color online) Two-port source wave amplitudes

FIG. 2. (a) Variation of condition number $\kappa$ of modal matrix $\boldsymbol{M}$ with axial spacing normalised by wavelength at 2 kHz, minimum indicated (b) Variation of condition number $\kappa$ for the ideal axial and azimuthal arrangement at the range of frequencies of interest

FIG. 3. (Color online) Schematic of measurement rig layout

FIG. 4. (Color online) Picture of measurement rig layout with test case fan and flow straightener installed

FIG. 5. (Color online) Non-dimensional pressure rise $\psi$ of test case fan as a function of flow coefficient $\phi$ with comparison to the performance measured in the ISO rig

FIG. 6. (Color online) Illustration of the azimuthal arrangement of the speakers where the speakers are driven with a phase shift between each to (preferentially) excite an azimuthal mode

FIG. 7. (Color online) Comparison between predictions and actual measurements for fan at the verification location on the (a) inlet side 1 (b) outlet side 2

FIG. 8. (Color online) Scattering matrix data at plane-wave frequencies for test case fan operating at its design point: (a) reflection (b) transmission

FIG. 9. (Color online) Variation of the predicted SPL (autospectral density) that would be measured under anechoic conditions for a duct cross-section on the inlet side

FIG. 10. (Color online) Variation of the predicted SPL (autospectral density) that would be measured under anechoic conditions for a duct cross-section on the outlet side